\documentclass[preprint,showpacs,preprintnumbers,amsmath,amssymb]{revtex4}
\usepackage{graphicx}
\usepackage{dcolumn}
\usepackage{bm}
\newcommand{\mbx}[1]{\mbox{\boldmath $#1$}}
\begin{document}
\preprint{cond-mat/0206286}
\title{Statistical Mechanics of \\
the Bayesian Image Restoration under Spatially Correlated Noise}
\author{Jun  Tsuzurugi${}^{\dagger}$${}^{\ddagger}$}
\email{juntuzu@brain.riken.go.jp}
\author{Masato  Okada${}^{\ddagger}$}%
\affiliation{
${}^{\dagger}$Nara Institute of Science and Technology,
Takayama-cho 8916-5, Ikoma-shi, Nara, 630-0101 Japan\\
${}^{\ddagger}$RIKEN Brain Science Institute
Hirosawa 2-1, Wakou-shi, Saitama, 351-0456 Japan
}
\date{\today}
\begin{abstract}
We investigated the use of the Bayesian inference to restore noise-degraded images  under conditions of spatially correlated noise. The generative statistical models used for the original image and the noise were assumed to obey multi-dimensional Gaussian distributions whose covariance matrices are translational invariant. We derived an exact description to be used as the expectation for the restored image by the Fourier transformation and restored an image distorted by spatially correlated noise by using a spatially uncorrelated noise model. 
We found that the resulting hyperparameter estimations for the minimum error and maximal posterior marginal criteria did not coincide when the generative probabilistic model and the model used for restoration were in different classes, while they did coincide when they were in the same class. 
\end{abstract}
\maketitle
\section{Introduction}
Recent research has shown that the Bayesian inference is a useful approach to the image restoration problem \cite{morina1}-\cite{inoue4}. However, previous research has usually assumed that the superimposed noise is not correlated between pixels \cite{morina1}-\cite{tanaka1}. Statistical mechanics provides a useful method for analyzing the image restoration problem \cite{nisi1},\cite{nisi2}. Still, when we take optical effects into consideration, it seems natural to take spatially correlated noise into account. 
We have investigated image restoration under the condition of spatially correlated noise. We considered a case in which both the original image and noise obey a Gaussian distribution composed of translational-symmetric matrices. 
We were able to diagonalize the Gaussian distribution by using Fourier transformation. 
While previous research was done using spatially uncorrelated noise, we considered 
the availability of a spatially uncorrelated noise model against spatially correlated noise. 

We considered tow methods for estimating the hyperparameters. 
The first is minimizing the mean squared error - the hyperparameters are set so that the mean squared error is minimized. However, minimizing the mean squared error is not applicable to the image restoration problem because we need the original image to practice this method.
The second method is maximizing the marginal likelihood 
\cite{tanaka2} - the hyperparameters are estimated by maximizing the marginal likelihood acquired from only the distorted image. When the noise probabilities for generating and restoring coincidence are the same, the results given by the two methods are the same. When the probabilities are different, we found that they are not. 
\section{Model}
The images we see in our daily life are always two-dimensional, although we assume a $d$-dimensional image for mathematical generalization, where the length of each side is $L$ and the total number of pixels, $N$, is $L^d$. 
The $\mbx{i}$-th pixel value, $\xi_{\mbx{i}}$, of an original image can be written in 
a multiple Gaussian distribution as
\begin{equation}
P({\mbx{\xi}})=\frac{1}{Z_{\rm{prior}}(\beta,h)}\exp
		\left[
			-H(\mbx{\xi},\beta,h)
		\right], \label{prior}
\end{equation}
\begin{equation}\label{hamil}
	H(\mbx{\xi},\beta,h)=\mbx{\xi}^{T}(\beta G + hI)\mbx{\xi},
\end{equation}
where $\beta$ and~$h$ are positive scalar values, $G$ is a matrix, and $I$ is the unit matrix. The partition function, $Z_{\rm{prior}}(\beta,h)$, is given by

\begin{equation}
Z_{\rm{prior}}(\beta,h)
	=	(2\pi)^{\frac{N}{2}}
		\left|
			2(\beta G + h I)
		\right|^{-\frac{1}{2}}.\label{Zprior}
\end{equation}
Note that Eq. (\ref{hamil}) is called the Hamiltonian in statistical mechanics.
In this paper, we argue that matrix $G$ is translational invariant and that element $G_{{\mbx{i}},{\mbx{j}}}$ is given by \cite{nisi2}

\begin{equation}
	G_{{\mbx{i}},{\mbx{j}}}
		= 2d \delta_{({\mbx{i}}-{\mbx{j}}),{\mbx{0}}}
		-\sum_{\mbx{\delta}}
			\delta_{({\mbx{i}}-{\mbx{j}}),-{\mbx{\delta}}}
		-\sum_{\mbx{\delta}}
			\delta_{({\mbx{i}}-{\mbx{j}}),{\mbx{\delta}}},\label{greenG}
\end{equation}
where the $d$-dimensional vector ${\mbx{\delta}}$ is

\begin{equation}
	{\mbx{\delta}} = (1,0,\cdots,0),(0,1,0,\cdots,0),\cdots,(0,\cdots,0,1).\label{pots}
\end{equation}
The $\delta_{\mbx{i},\mbx{j}}$ in Eq. (\ref{greenG}) is the Kronecker delta, which is assumed to be

\begin{equation}
	\delta_{\mbx{i},\mbx{j}}=
	\left\{
		\begin{array}{rl}
		1 & \mbox{(\(\mbx{i}=\mbx{j}\))} \\
		0 & \mbox{(\(\mbx{i}\neq\mbx{j}\)).}
		\end{array}
	\right.
\end{equation}
Equation (\ref{greenG}) means that interactions occur only between nearest-neighbor pixels.
If $\beta$ takes a large value, i.e., the interaction of the first term has a large effect, the neighboring pixels tend to take the same value. When $h$ is large, the absolute value of the pixels tends to be small. Furthermore, we use a periodic boundary condition, 

\begin{equation}
	\mbx{\xi}_{\mbx{i}} = \mbx{\xi}_{\mbx{i}+L\mbx{\delta}}.
\end{equation}
The distorted image, ${\mbx{\tau}}= \{\tau_{\mbx{i}}\}$, is generated according to the following conditional probability, which obeys a Gaussian distribution: 
\begin{equation}
	P_{\rm{out}}({\mbx{\tau}}|{\mbx{\xi}})=
		\frac{1}{Z_{\rm{noise}}}
		\exp\left[
			-\frac{1}{2}
				({\mbx{\tau}}-{\mbx{\xi}})^T
				R^{-1}
				({\mbx{\tau}}-{\mbx{\xi}})
		\right],\label{noize}
\end{equation}
\begin{equation}
Z_{\rm{noise}}(R)=(2\pi)^{\frac{N}{2}}|R|^{\frac{1}{2}},\label{Znoize}
\end{equation}
where $R$ is a translational invariant covariance matrix.
In this paper, we consider the following spatially correlated noise:

\begin{equation}\label{soukan}
	R_{{\mbx{i}},{\mbx{j}}}
		= (1-a)b^2 \delta_{{\mbx{i}},{\mbx{j}}}
			+a b^2 \exp\left[-\frac{({\mbx{i}}-{\mbx{j}})^2}
						{\kappa^2}
			\right],
\end{equation}
$0\leq a \leq 1$.
The noise obeys an identical independent Gaussian distribution with mean $0$ and variance $b^2$, when $a=0$. 

A general strategy commonly used in image restoration is to apply the Bayes formulation to the posterior probability. Here, $\mbx{\sigma}$ is a restoration image  based on the Bayes formulation. When a distorted image, $\mbx{\tau}$, is given, one can use the formulation to calculate the restored image, $\mbx{\sigma}$:

	\begin{equation}
	P(\mbx{\sigma}|\mbx{\tau})
	=
		\frac{
			P_{out}(\mbx{\tau}|\mbx{\sigma})P(\mbx{\sigma})
		}
		{
			\int d\sigma
				P_{out}(\mbx{\tau}|\mbx{\sigma})P(\mbx{\sigma})
		}
	=\frac
		{
			\exp \left[ -H_{eff} \right]
		}
		{
			\int d \sigma
			\exp{\left[ -H_{eff} \right]}
		},\label{Bayes}
\end{equation}
where

\begin{equation}
H_{eff} = 
			\mbx{\sigma}^{T}(\beta G + hI)\mbx{\sigma}
			+\frac{1}{2}
				({\mbx{\tau}}-{\mbx{\sigma}})^T
				R^{-1}
				({\mbx{\tau}}-{\mbx{\sigma}}),\label{yuukouA}
\end{equation}
is an effective Hamiltonian.
\section{Theory}
\subsection{Expectation of the restored image}
Since the covariance matrices given by Eqs. (\ref{greenG}) and (\ref{soukan}) are translational invariant, they can be diagonalized using the discrete Fourier transformation.

The discrete Fourier transformation is
\begin{equation}\label{Fourier}
	\tilde{\xi}_{\mbx{k}} = \frac{1}{\sqrt{N}}
					\sum_{\mbx{j}} \xi_{\mbx{j}}
						e^{-i{\mbx{k\cdot j}}} ,
\end{equation}
and the inverse Fourier transformation is

\begin{equation}\label{gaku}
	\xi_{\mbx{j}} = \frac{1}{\sqrt{N}}
					\sum_{\mbx{k}} \tilde{\xi}_{\mbx{k}}
						e^{i{\mbx{k\cdot j}}} ,
\end{equation}
where $i$ is the imaginary unit, and $\mbx{k}$ is a $d$-dimensional vector with the same as $\mbx{j}$, and the degree of freedom is $L^d$.
\begin{equation}
	\sum_{\mbx{j}}=L^d=N,~~~\sum_{\mbx{k}}=L^d=N.
\end{equation}
Moreover, each component of ${\mbx{k}}$ takes the value
\begin{equation}
	0,\frac{2}{L}\pi, \frac{4}{L}\pi,\cdots,\frac{2(L-1)}{L}\pi.
\end{equation}
We can diagonalize Eq. (\ref{prior}) by using the Fourier representation:
\begin{equation}
	P({\mbx{\xi}})=
	\frac{1}{Z_{\rm{prior}}(\beta,h)}\exp
		\left[
				-\sum_{\mbx{k}}
				(\beta \tilde{G}_{\mbx{k}} + h)
				\tilde{\xi}_{\mbx{k}}\tilde{\xi}_{-\mbx{k}}
		\right],
	\label{priorF}
\end{equation}
\begin{equation}\label{green}
	\tilde{G}_{\mbx{k}}
		= \sum_{\mbx{\delta}}
			[2-2\cos({\mbx{k}}\cdot{\mbx{\delta}})].
\end{equation}
Furthermore, we diagonalize $R_{\mbx{i},\mbx{j}}$ in Eq. (\ref{soukan}):

\begin{equation}\label{soukanF}
	\tilde{R}_{\mbx{k}}
	=(1-a)b^2+ab^2\sum_{\mbx{l}}
		e^{-\frac{{\mbx{l}}^2}
						{\kappa^2}
		}
				\cos({\mbx{k}}\cdot{\mbx{l}}),
\end{equation}
where the range of $l$, which is a component of vector $\mbx{l}$, is given by

\begin{equation}
-(L-1)\leq l \leq L-1.
\end{equation}
We can execute the Fourier transformation on the inside of $\exp$ in Eq. (\ref{noize}) 
in the same way as in Eq. (\ref{prior}).

\begin{eqnarray}
	& &P_{out}({\mbx{\tau}}|{\mbx{\xi}})
	=\frac{1}{Z_{\rm{noise}}} \nonumber \\
	& &\times \exp\left[
			-\frac{1}{2}\sum_{\mbx{k}}
				\tilde{R}_{\mbx{k}}^{-1}
				(\tilde{\tau}_{\mbx{k}}-\tilde{\xi}_{\mbx{k}})
				(\tilde{\tau}_{-{\mbx{k}}}-\tilde{\xi}_{-{\mbx{k}}})
		\right]. \label{Pout}
\end{eqnarray}
We define $\left< \cdot \right>$ as denoting the thermal average 
based on the Bayes formula for the posterior probability $P(\mbx{\sigma}|\mbx{\tau})$ in Eq. (\ref{Bayes}). The expectation of the restored image, $\sigma_{\mbx{j}}$, is thus assumed to be
\begin{equation}\label{repair}
	\left<\sigma_{\mbx{j}}\right>
		=
		\frac{1}{\sqrt{N}}
			\sum_{\mbx{k}}
			\left<
				\tilde{\sigma}_{\mbx{k}}
			\right>
			e^{i{\mbx{k}}\cdot{\mbx{j}}}.
\end{equation}

Using Eq. (\ref{Bayes}), the expectation of the Fourier component, $\sigma_{\mbx{k}}$, can be written as
\begin{equation}
		\left< \tilde{\sigma}_{\mbx{k}} \right>
	=\int \prod_{\mbx{k}'} d\tilde{\sigma}_{\mbx{k}'}
		\tilde{\sigma}_{\mbx{k}}
		P(\mbx{\sigma}|\mbx{\tau})
	=\frac
		{
			\int \prod_{\mbx{k}'} d\tilde{\sigma}_{\mbx{k}'}
				\tilde{\sigma}_{\mbx{k}} e^{-\hat{\tilde{H}}_{eff}}
		}
		{
			\int \prod_{\mbx{k}'} d\tilde{\sigma}_{\mbx{k}'}
				e^{-\hat{\tilde{H}}_{eff}}
		},\label{kakkonai}
\end{equation}
where $\hat{\tilde{H}}_{eff}$ is obtained by applying the Fourier transformation to Eq. (\ref{yuukouA}) as follows.
\begin{eqnarray}
	\hat{\tilde{H}}_{eff} &=&
		\sum_{\mbx{k}}
			({\hat{\beta}} \tilde{G}_{\mbx{k}} + \hat{h})
				\tilde{\sigma}_{\mbx{k}} \tilde{\sigma}_{\mbx{-k}}
				\nonumber \\
		&+&\sum_{\mbx{k}}\frac{1}{2\hat{\tilde{R}}_{\mbx{k}}}
				(\tilde{\tau}_{\mbx{k}}-\tilde{\sigma}_{\mbx{k}})
				(\tilde{\tau}_{-{\mbx{k}}}-\tilde{\sigma}_{-{\mbx{k}}}).
					\label{yuukou}
\end{eqnarray}
$\beta$, $h$, and $\tilde{R}_{\mbx{k}}$ are unknown adjustable parameters that determine the properties of the original image and noise. One of our major focus here is how to estimate these hyperparameters precisely. 

To proceed further, we have to assume some explicit form of the source prior and noise posterior formations, which are used when we restore the image. In this paper, we define both of these probability formations as the same formation except that they have different hyperparameters. We define the hyperparameters that correspond to $\beta, h,$, and $\tilde{R}_{\mbx{k}}$ as $\hat{\beta}$, $\hat{h}$, and $\hat{\tilde{R}}_{\mbx{k}}$, respectively. If $\hat{\tilde{R}}_{\mbx{k}}$ is independent of $\mbx{k}$, the noise is spatially uncorrelated.

Substituting
\begin{equation}
	\hat{\tilde{A}}_{\mbx{k}}
	=\hat{\beta} \tilde{G}_{\mbx{k}}
		+\hat{h} +\frac{1}{2\hat{\tilde{R}}_{\mbx{k}}},~~~~~~~~~~~
	\hat{\tilde{B}}_{\mbx{k}}
	=\frac{1}{2\hat{\tilde{R}}_{\mbx{k}}}
\end{equation}
\label{okikae}
into Eqs. (\ref{kakkonai}) and (\ref{yuukou}), we get
\begin{equation}
	\left< \tilde{\sigma}_{\mbx{k}} \right>
	=\frac{\hat{\tilde{B}}_{\mbx{k}}}{\hat{\tilde{A}}_{\mbx{k}}}
	\tilde{\tau}_{\mbx{k}}\label{rep}.
\end{equation}
Consequently, $\left< \sigma_{\mbx{j}} \right>$ in Eq. (\ref{repair}) is given by
\begin{equation}
	\left< \sigma_{\mbx{j}} \right>=
		\frac{1}{N}
		\sum_{\mbx{k}}\sum_{\mbx{i}}
		\frac{
			\tau_{\mbx{i}}
			\frac{1}{2\hat{\tilde{R}}_{\mbx{k}}}
			\cos[
				{{\mbx{k}}\cdot({\mbx{j}}-{\mbx{i}})}
			]
		}
		{
			\hat{\beta} \tilde{G}_{\mbx{k}}+\hat{h}
				+\frac{1}{2\hat{\tilde{R}}_{\mbx{k}}}
		}.\label{E1kaiseki}
\end{equation}
In this paper, we regard Eq. (\ref{E1kaiseki}) as representing the restored image.
\subsection{Minimization of mean squared error}
In this subsection, we estimate hyperparameters $\hat{\beta}, \hat{h}$, and $\hat{\tilde{R}}_{\mbx{k}}$ by using the criterion defined by Eq. (\ref{E1kaiseki}) for minimizing the mean squared error between the original and restored images. 
The expectation of this mean squared error, $E_1$, is represented by
\begin{equation}
	E_1=
		\left\|
		\frac{1}{N}\sum_{\mbx{j}}
			\left(
				\xi_{\mbx{j}}
				-\left<
					\sigma_{\mbx{j}}
				\right>
			\right)^2
	\right\|,\label{E1}
\end{equation}
where $\left\| \cdot \right\|$
denotes the data average of a simultaneous distribution\\
	$P(\mbx{\tau},\mbx{\xi})=
	P_{out}({\mbx{\tau}}|{\mbx{\xi}})
	P_s({\mbx{\xi}})$.

Applying the Fourier transformation to Eq. (\ref{E1}), we can derive the next representation.
\begin{equation}
		E_1=
		\frac{1}{N}\sum_{\mbx{k}}
		\left\|
			\Big(
				\tilde{\xi}_{\mbx{k}}
				-\left<\tilde{\sigma}_{\mbx{k}}\right>
			\Big)
			\Big(
				\tilde{\xi}_{-{\mbx{k}}}
				-\left<\tilde{\sigma}_{-{\mbx{k}}}\right>
			\Big)
		\right\|. \label{EE1}
\end{equation}
Since $P(\mbx{\tau},\mbx{\xi})$ is diagonalized by the Fourier transformation, we can easily calculate each ${\mbx{k}}$ as follows,
\begin{eqnarray}
	& &\left\|
		\Big(
			\tilde{\xi}_{\mbx{k}}
			-\left<
				\tilde{\sigma}_{\mbx{k}}
			\right>
		\Big)
		\Big(
			\tilde{\xi}_{-\mbx{k}}
			-\left<
				\tilde{\sigma}_{-\mbx{k}}
			\right>
		\Big)
		\right\| \\
	&=&	\frac{1}{Z}
		\int\int
			d\tilde{\xi}_{\mbx{k}}
			d\tilde{\tau}_{\mbx{k}}
			\left|
				\tilde{\xi}_{\mbx{k}}
				-\frac{\hat{\tilde{B}}_{\mbx{k}}}{\hat{\tilde{A}}_{\mbx{k}}}
				\tilde{\tau}_{\mbx{k}}
			\right|^2\nonumber \\
	& &\times\exp
		\Bigg[
			-A_{\mbx{k}}
				\left|
					\tilde{\xi}_{\mbx{k}}
					-\frac{\tilde{B}_{\mbx{k}}}{\tilde{A}_{\mbx{k}}}
					\tilde{\tau}_{\mbx{k}}
				\right|^2\nonumber \\
	& &~~~~~~~~~~-
				\left(
					\tilde{B}_{\mbx{k}}
					-\frac{\tilde{B}_{\mbx{k}}^2}{\tilde{A}_{\mbx{k}}}
				\right)
				\left|
					\tilde{\tau}_{\mbx{k}}
				\right|^2
		\Bigg]\\
	&=&\frac{1}{2\tilde{A}_{\mbx{k}}}+
		\left(
			\frac{\tilde{B}_{\mbx{k}}}{\tilde{A}_{\mbx{k}}}
			-\frac{\hat{\tilde{B}}_{\mbx{k}}}{\hat{\tilde{A}}_{\mbx{k}}}
		\right)^2
		\frac{1}{
				2\tilde{B}_{\mbx{k}}
				-\frac{2\tilde{B}_{\mbx{k}}^2}{\tilde{A}_{\mbx{k}}}
		},\label{11}
\end{eqnarray}
where $\tilde{A}_{\mbx{k}}$, $\tilde{B}_{\mbx{k}}$, and $Z$ substitute for
\begin{equation}
	\tilde{A}_{\mbx{k}}=\beta \tilde{G}_{\mbx{k}} + h
		+\frac{1}{2\tilde{R}_{\mbx{k}}},~~~~~
	\tilde{B}_{\mbx{k}} = \frac{1}{2\tilde{R}_{\mbx{k}}},
\end{equation}
\begin{eqnarray}
	Z
	&=&	\int\int
			d\tilde{\xi}_{\mbx{k}}
			d\tilde{\tau}_{\mbx{k}}
		\exp
		\Bigg[
			-\tilde{A}_{\mbx{k}}
				\left|
					\tilde{\xi}_{\mbx{k}}
					-\frac{\tilde{B}_{\mbx{k}}}{\tilde{A}_{\mbx{k}}}
					\tilde{\tau}_{\mbx{k}}
				\right|^2\nonumber \\
	& &~~~~~~~~~~~~~~~~~~~~~~~~~-
				\left(
					\tilde{B}_{\mbx{k}}
					-\frac{\tilde{B}_{\mbx{k}}^2}{\tilde{A}_{\mbx{k}}}
				\right)
				\left|
					\tilde{\tau}_{\mbx{k}}
				\right|^2
		\Bigg].
\end{eqnarray}
Therefore, the mean squared error is represented by
\begin{eqnarray}
	& &E_1= 
		\frac{1}{2N}\sum_{\mbx{k}}
	\Biggl[
		\frac{1}{\beta \tilde{G}_{\mbx{k}}+h +\frac{1}{2\tilde{R}_{\mbx{k}}}}
		\nonumber \\
	& &+\left(
			\frac{\frac{1}{2\tilde{R}_{\mbx{k}}}}
				{\beta \tilde{G}_{\mbx{k}} +h +\frac{1}{2\tilde{R}_{\mbx{k}}}}
			-\frac{\frac{1}{2\hat{\tilde{R}}_{\mbx{k}}}}
				{\hat{\beta} \tilde{G}_{\mbx{k}}+\hat{h}
				+\frac{1}{2\hat{\tilde{R}}_{\mbx{k}}}}
		\right)^2\nonumber \\
	& &
		~~~~~~~~~~~~~~\times\frac{
				\beta \tilde{G}_{\mbx{k}}
				+h +\frac{1}{2\tilde{R}_{\mbx{k}}}
		}
		{
			\frac{1}{2\tilde{R}_{\mbx{k}}}
			\left(
				\beta \tilde{G}_{\mbx{k}}+h
			\right)
		}
		\Biggr].\label{Eend}
\end{eqnarray}
When the conditions 
\begin{equation}
\hat{\beta}=\beta, ~\hat{h}=h, {\rm and} ~\hat{{\tilde{R}}}_{\mbx{k}} = {\tilde{R}}_{\mbx{k}},
~\label{saitekipara}
\end{equation}
hold, the mean squared error takes the following minimum:
\begin{equation}
	E_{1_{min}}
	= \frac{1}{2N}\sum_{\mbx{k}}
	\left[
		\frac{1}{\beta \tilde{G}_{\mbx{k}}+h +\frac{1}{2\tilde{R}_{\mbx{k}}}}
	\right]\label{saitei},
\end{equation}
More precisely, 
when the ratio $\hat{\beta}: \hat{h}:\hat{\tilde{R}}_{\mbx{k}}$
is equal to the ratio ${\beta}:{h}:{\tilde{R}}_{\mbx{k}}$, 
$E_1$ is the minimum given by Eq. (\ref{saitei}). 
Note that this equation represents the limit of the restoration.

Likewise, we can derive the mean squared error, $E_2$, between the original image and the distorted image as being represented by
\begin{eqnarray}
	E_2&=&\left\|
		\frac{1}{N}\sum_{\mbx{j}}
			\left(
				\tilde{\xi}_{\mbx{j}}
				-\tilde{\tau}_{\mbx{j}}
			\right)^2
	\right\|\label{E2} \\
	&=&\frac{1}{N}
		\sum_{\mbx{k}}{\tilde{R}_{\mbx{k}}}
	=\frac{1}{N}\sum_{\mbx{i}}\tilde{R}_{\mbx{i},\mbx{i}}=b^2
	\label{kaisekiE2}
\end{eqnarray}
$E_2$ depends only on the diagonal element of the noise covariance matrix.
\subsection{Maximization of the marginal likelihood}
The minimized mean squared error criterion generally cannot be used since the unknown original image itself is needed to evaluate the squared error. If we already had the original image, there would be no need to restore the distorted image. Hence, in this paper, we argue for using the maximization of the marginal likelihood. 

As shown in Eqs. (\ref{prior}) and (\ref{noize}), we already know the source prior and noise probabilities. Using both probabilities, we can derive $P({\mbx{\tau}})$:

\begin{equation}
P({\mbx{\tau}})=\int d{\mbx{S}}
	P_{\rm{out}} (\mbx{\tau}|\mbx{S};\hat{R})P(\mbx{S};\hat{\beta},\hat{h}).\label{syuhen1}
\end{equation}
$P({\mbx{\tau}})$ is called the marginal likelihood.

Maximization of the marginal likelihood is used to set the hyperparameters, $\hat{\beta}, \hat{h}$, and $\hat{R}$, to use to obtain the maximum value of $P(\mbx{\tau})$ for distorted image $\mbx{\tau}$.

With Eqs. (\ref{Zprior}) and (\ref{Znoize}), the marginal likelihood is given by
\begin{equation}
P({\mbx{\tau}})=\frac{Z_{\rm{posterior}}(\hat{\beta},\hat{h},\hat{R})}
			{Z_{\rm{noise}}(\hat{R})Z_{\rm{prior}}(\hat{\beta},\hat{h})},
\end{equation}
where
\begin{eqnarray}
	& &Z_{\rm{posterior}}(\hat{\beta},\hat{h},\hat{R})
	=\prod_{\mbx{k}}\int d\tilde{S}_{\mbx{k}}\exp(-\hat{\tilde{H}}_{eff})\\
	&=&\pi^{\frac{N}{2}}
			\prod_{\mbx{k}}{\frac{1}{\sqrt{\hat{\tilde{A}}_{\mbx{k}}}}}
		\exp
				\left[
					-(\hat{\tilde{B}}_{\mbx{k}}
					 - \frac{\hat{\tilde{B}}^2_{\mbx{k}}}{\hat{\tilde{A}}_{\mbx{k}}})
						\left|\tilde{\tau}_{\mbx{k}}\right|^2
				\right].\label{Zpost}
\end{eqnarray}

Because $\ln$ is a monotonically increasing function that maximizes the logarithmic marginal likelihood, maximizing $\ln P(\mbx{\tau})$ is equivalent to maximizing $P(\mbx{\tau})$.

The logarithmic marginal likelihood $\ln P(\mbx{\tau})$ can be written as,
\begin{eqnarray}
	& &\ln\left(P({\mbx{\tau}})\right)
	=\ln
		\left(
			\frac{Z_{\rm{posterior}}(\hat{\beta},\hat{h},\hat{R})}
			{Z_{\rm{noise}}(\hat{R})Z_{\rm{prior}}(\hat{\beta},\hat{h})}
		\right)\nonumber \\
	&=&-\frac{1}{2}\sum_{\mbx{k}}\ln
		\left(
			\hat{\beta} \tilde{G}_{\mbx{k}}+\hat{h}+\frac{1}{2\hat{\tilde{R}}_{\mbx{k}}}
		\right)
		-\frac{N}{2}\ln(2\pi)\nonumber \\
	& &-\frac{1}{2}\sum_{\mbx{k}}\ln 
	\left(\hat{\tilde{R}}_{\mbx{k}}\right)
	+\frac{1}{2}\sum_{\mbx{k}}\ln
		\left(
			\hat{\beta} G_{\mbx{k}} + \hat{h}
		\right)\nonumber \\
	& &-\sum_{\mbx{k}}
		\frac{
			\left(\hat{\beta} \tilde{G}_{\mbx{k}}
			 + \hat{h}\right)\frac{1}{2\hat{\tilde{R}}_{\mbx{k}}}
		}
		{
			\hat{\beta} \tilde{G}_{\mbx{k}} + \hat{h} +\frac{1}{2\hat{\tilde{R}}_{\mbx{k}}}
		}
		\left|\tilde{\tau}_{\mbx{k}}\right|^2.\label{syuhen2}
\end{eqnarray}
\section{result}
In this section, we consider a method to restore an image distorted by spatially correlated noise by means of the spatially uncorrelated noise model. To do this, we replace $R$ as one of the hyperparameters:
\begin{equation}
\hat{R}_{\mbx{i},\mbx{j}}=\hat{r}\delta_{\mbx{i},\mbx{j}},
~~~~~~~~~~~\hat{\tilde{R}}_{\mbx{k}}=\hat{r}.\label{kinjiF}
\end{equation}
We can now write Eq. (\ref{E1kaiseki}) as
\begin{equation}
		\left< \sigma_{\mbx{j}}^{app} \right>
	=\frac{1}{N}
		\sum_{\mbx{k}}\sum_{\mbx{i}}
		\frac{
			\tau_{\mbx{i}}
			\frac{1}{2r}
			\cos[
				{{\mbx{k}}\cdot({\mbx{j}}-{\mbx{i}})}
			]
		}
		{
			\hat{\beta} \tilde{G}_{\mbx{k}}+\hat{h}
				+\frac{1}{2\hat{r}}
		}.\label{junsaiteki}
\end{equation}
\begin{figure}[ht]
	\centering
	\includegraphics[width=8.5cm,clip]{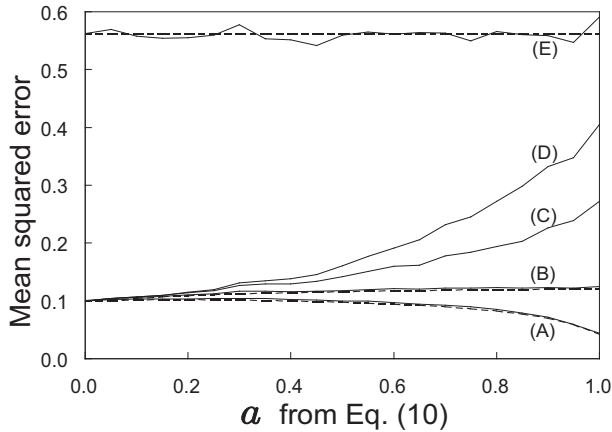}
\caption{
Mean squared error between original and each restored image. Horizontal axis denotes $a$ from Eq. (\ref{soukan}); vertical axis denotes mean squared error. (A) Optimum decode given by Eq. (\ref{saitei}). (B) Restored using criterion of minimum restored error. (C) Restored using $\hat{r}$ estimated by maximizing marginal likelihood. (D) Restored using $\hat{\beta}, \hat{h}$, and $\hat{r}$, all estimated by maximizing marginal likelihood. (E) Error before decoding given by Eq. (\ref{kaisekiE2}).
}
\label{kekkagraf}
\end{figure}
\begin{figure}[ht]
	\centering
	\includegraphics[width=8cm,clip]{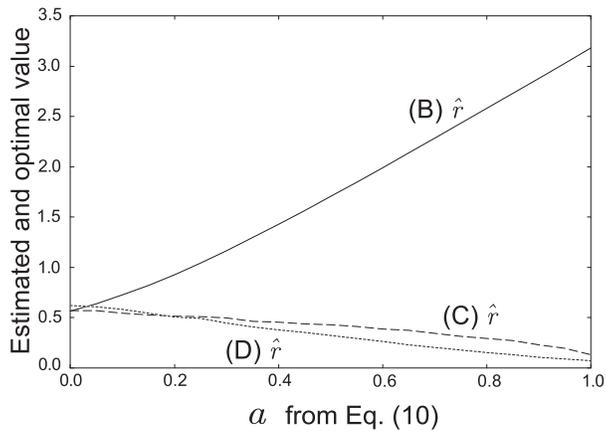}
\caption{
	Estimated $\hat{r}$. 
	Horizontal axis denotes $a$ from Eq. (\ref{soukan}); 
	vertical axis denotes average estimated 
	$\hat{r}$ 
	obtained by maximizing marginal likelihood 
	or optimal $\hat{r}$ 
	obtained using criterion of minimum restored error. 
	Solid line (B) denotes $\hat{r}$ used in Fig. \ref{kekkagraf}(B), 
	dashed line (C) denotes $\hat{r}$ used in Fig. \ref{kekkagraf}(C), 
	and dotted line (D) denotes $\hat{r}$ used in Fig. \ref{kekkagraf}(D).
}
\label{reps}
\end{figure}
\begin{figure}[ht]
	\centering
	\includegraphics[width=8cm,clip]{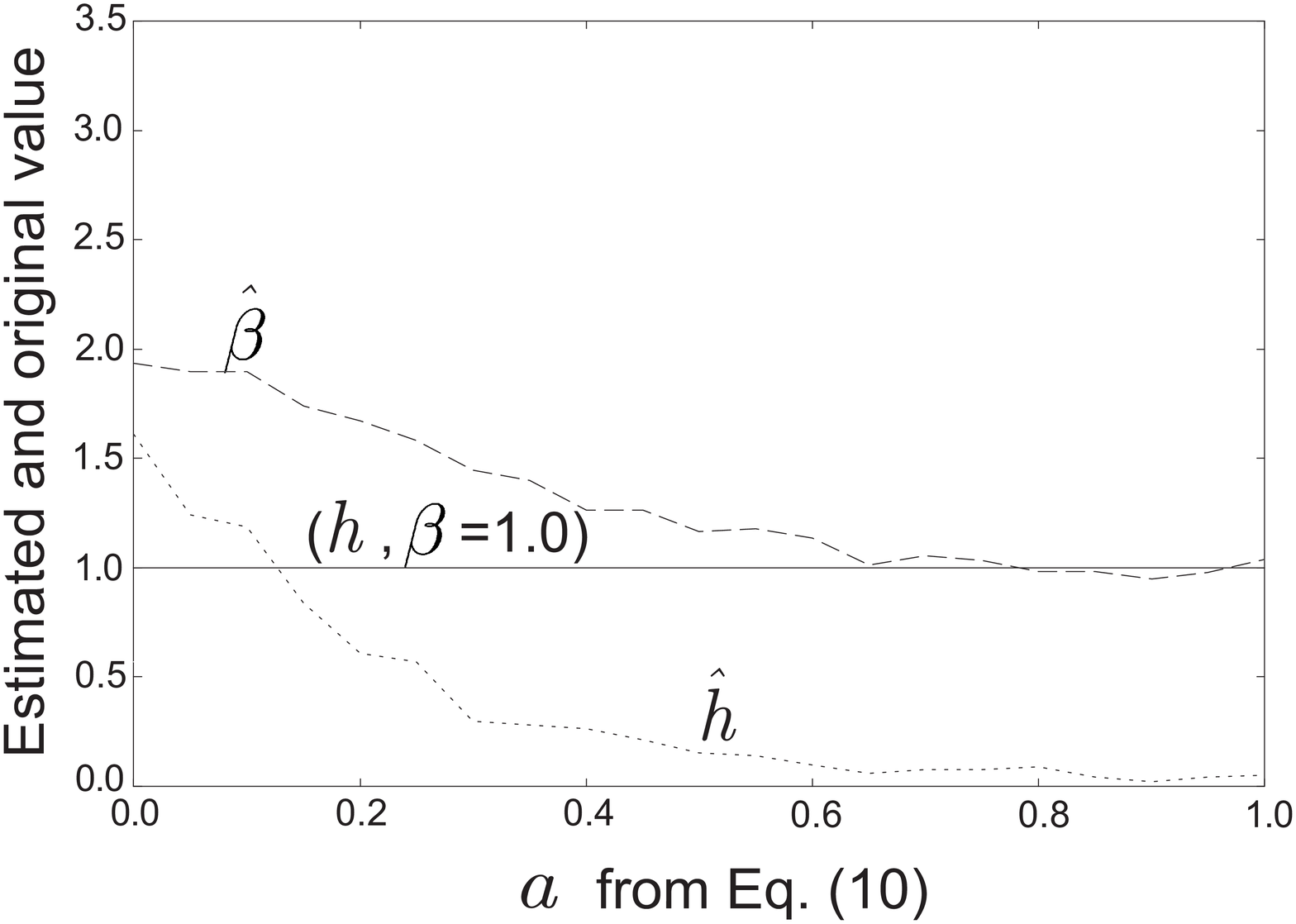}
\caption{
	Estimated $\hat{h}$ (dotted line) and $\hat{\beta}$ (dashed line). 
	Both values were averaged with respect to 100 artificially generated images. 
	Horizontal axis denotes $a$ from Eq. (\ref{soukan}); 
	vertical axis denotes average estimated $\hat{h}$ or $\hat{\beta}$
	or actual $h$ or $\beta$. 
	Solid line denotes actual $h (=\beta=1.0$).
}
\label{hbeps}
\end{figure}
Figure \ref{kekkagraf} shows the analytical 
and simulated mean squared errors between the original image and each restored image, 
Fig. \ref{reps} shows the estimated $\hat{r}$, 
and Fig. \ref{hbeps} show the estimated parameters $\hat{h}$ and $\hat{\beta}$. 
We used 100 artificially generated two-dimensional $N=16^2$ images with hyperparameters $\beta=1.0$, $h=1.0$, $b=0.75$, and $\kappa=3.0$. 
The horizontal axes of these three figures denote $a$ from Eq. (\ref{soukan}); 
$a$ represents the magnitude of the spatial correlation of the noise ($a=0$ implies no spatial correlation). 

\subsection{Minimization of mean squared error}
Curve (E) in Fig. \ref{kekkagraf} represents error $E_2$ given by Eq. (\ref{kaisekiE2}), that is, the mean squared error between the original and distorted images. (The dashed line represents the analytical error, and the solid line represents the simulated results.) Because the numbers of pixels ($N=16^2$) and samples ($100$) were finite, the simulated results have some fluctuation. This fluctuation tends to asymptotically disappear as the numbers are increased.

Curve (A) in Fig. \ref{kekkagraf} represents minimum error $E_{1_{min}}$ given by Eq. (\ref{saitei}), that is, the mean squared error between the original and distorted images restored using the spatially correlated noise model with the best parameters. Hence, $E_{1_{min}}$ implies the limits of image restoration-no further error reduction is possible.
Curve (B) in Fig. \ref{kekkagraf} represents the error in an image restored using the uncorrelated noise model. The restoration was done using hyperparameter $\hat{r}$ from Eq. (\ref{E1}), which was acquired minimizing the mean squared error. (The dashed line represents the analytical results, and the solid line represents the simulated results.)
Curve (B) in Fig. \ref{reps} shows the optimal values of $\hat{r}$ 
by minimizing the mean squared error.

Note that when $a=0.0$, curve (A) in Fig. \ref{kekkagraf} is equal to curve (B).
Curve (B) in Fig. \ref{kekkagraf} shows the optimum performance 
when using the minimization of the mean squared error as the criterion. 
However, as mentioned, the hyperparameters cannot be estimated in practice by minimizing the mean squared error. 
Therefore, we will discuss how the error can be reduced 
by maximizing the logarithmic marginal likelihood in the next subsection.

\subsection{Maximization of the marginal likelihood}
Substituting Eq. (\ref{syuhen2}) for Eq. (\ref{kinjiF}), we get
\begin{eqnarray}
	\ln\left(P_{app}({\mbx{\tau}})\right)
		&=&-\frac{1}{2}\sum_{\mbx{k}}\ln
		\left(
			\hat{\beta} \tilde{G}_{\mbx{k}}+\hat{h}+\frac{1}{2\hat{r}}
		\right)-\frac{N}{2}\ln(2\pi)\nonumber \\
	& &	-\frac{N}{2}\ln \left(\hat{r}\right)
		+\frac{1}{2}\sum_{\mbx{k}}\ln
		\left(
			\hat{\beta} \tilde{G}_{\mbx{k}} + \hat{h}
		\right)\nonumber \\
	& &-\sum_{\mbx{k}}
		\frac{
			\left(\hat{\beta} \tilde{G}_{\mbx{k}} + \hat{h}\right)\frac{1}{2\hat{r}}
		}
		{
			\hat{\beta} \tilde{G}_{\mbx{k}} + \hat{h} +\frac{1}{2\hat{r}}
		}
		\left|\tilde{\tau}_{\mbx{k}}\right|^2.\label{syuhenapp}
\end{eqnarray}
We can derive the conditions necessary for the extremal $P_{app}({\mbx{\tau}})$, and we can represent them using the following nonlinear simultaneous equations.
\begin{eqnarray}
	\frac{1}{\hat{\beta}}
	&=&	\left(
			\sum_{\mbx{k}}
				\frac{
					\tilde{G}_{\mbx{k}}
				}
				{
					\tilde{G}_{\mbx{k}}+\frac{\hat{h}}{\hat{\beta}}
				}
		\right)^{-1}
		\Bigg[
			\sum_{\mbx{k}}\frac{\tilde{G}_{\mbx{k}}}
						{
							\hat{\beta} \tilde{G}_{\mbx{k}} + \hat{h} + \frac{1}{2\hat{r}}
						}\nonumber \\
	& &	~~~~~~+\sum_{\mbx{k}}\left|\tau_{\mbx{k}}\right|^2
			\frac{\tilde{G}_{\mbx{k}}}{2\hat{r}^2
				\left(
					\hat{\beta} \tilde{G}_{\mbx{k}} + \hat{h} + \frac{1}{2\hat{r}}
				\right)^2}
		\Bigg],\label{b}
\end{eqnarray}
\begin{eqnarray}
	\hat{r}
	&=&	\frac{1}{2N}
			\sum_{\mbx{k}}\frac{\tilde{G}_{\mbx{k}}}
						{
							\hat{\beta} \tilde{G}_{\mbx{k}} + \hat{h} + \frac{1}{2\hat{r}}
						}\nonumber \\
	& &~~~~~~~~~+\frac{1}{N}
		\sum_{\mbx{k}}\left|\tilde{\tau}_{\mbx{k}}\right|^2
			\frac{(\hat{\beta} \tilde{G}_{\mbx{k}} + \hat{h})^2}
				{
				\left(
					\hat{\beta} \tilde{G}_{\mbx{k}} + \hat{h} + \frac{1}{2\hat{r}}
				\right)^2},\label{q}
\end{eqnarray}
\begin{eqnarray}
	\frac{1}{\hat{h}}
	&=&	\left(
			\sum_{\mbx{k}}
				\frac{
						1
				}
				{
					1+\frac{\hat{\beta} \tilde{G}_{\mbx{k}}}{h}
				}
		\right)^{-1}
		\Bigg[
			\sum_{\mbx{k}}\frac{1}
						{
							\hat{\beta} \tilde{G}_{\mbx{k}} + \hat{h} + \frac{1}{2\hat{r}}
						}\nonumber \\
	& &~~~~~~+\sum_{\mbx{k}}\left|\tau_{\mbx{k}}\right|^2
			\frac{1}{2\hat{r}^2
				\left(
					\hat{\beta} \tilde{G}_{\mbx{k}} + \hat{h} + \frac{1}{2\hat{r}}
				\right)^2}
		\Bigg].\label{h}
\end{eqnarray}
We can then determine the convergence point by means of an iterative calculation.

We next determined the unknown hyperparameter, $\hat{r}$, by maximizing the marginal likelihood, assuming that $\hat{\beta} = \beta$ and $\hat{h}=h$. Thus, we restricted ourselves to the case where the generation model of the image was already known, as was $\beta$ and $h$.
Curve (C) in Fig. \ref{kekkagraf} shows the mean squared error when the image was 
restored using the estimated $\hat{r}$. 
Curve (C) in Fig. \ref{reps} shows the estimated $\hat{r}$.

Note that the difference between
curve (B) in Fig. \ref{reps}, the limit of the restoration,
and curve (C), obtained by minimizing the mean squared error 
increased with the noise correlation parameter, $a$. 
Therefore, the error obtained by maximizing the marginal likelihood 
(curve (C) in Fig. \ref{kekkagraf}) increases with the noise correlation parameter, $a$.
This implies that without knowing the noise model, estimating the hyperparameters by maximizing the marginal likelihood does not work well. That is, the conventional uncorrelated noise model cannot cope with spatially correlated noise.

To confirm this, we used maximization of the marginal likelihood to estimate 
$\hat{\beta}$ and $\hat{h}$-the hyperparameters of the image-generation probability-as well as $\hat{r}$. 
Curve (D) in Fig. \ref{kekkagraf} shows the mean squared error estimated using hyperparameters $\hat{\beta}$, $\hat{h}$, and $\hat{r}$. 
Curve (D) in Fig. \ref{reps} shows the obtained $\hat{r}$, and Fig. \ref{hbeps} shows the 
obtained $\hat{\beta}$ and $\hat{h}$. 
The error shown by curve (D) in Fig. \ref{kekkagraf} is much greater than that shown by curve (C). 

\begin{figure}[ht]
	\centering
	\includegraphics[width=8cm,clip]{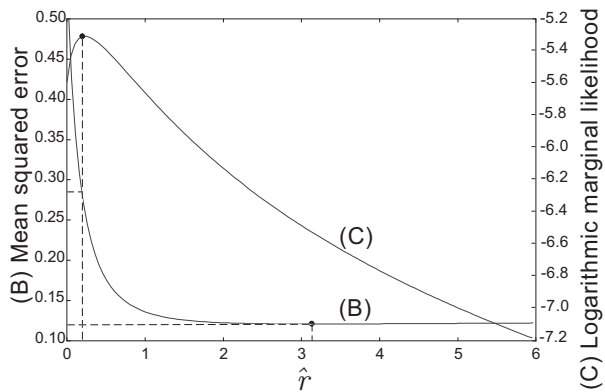}
\caption{
	Mean squared error and marginal likelihood.
	Horizontal axis denotes hyperparameter $\hat{r}$;
	left vertical axis denotes mean squared error (B) given by Eq.(\ref{Eend}); 
	right vertical axis denotes logarithmic marginal likelihood (C)
	given by Eq.(\ref{syuhenapp}).
	The parameters were 
	$a=1.0$, $\beta=1.0$, $h = 1.0$, and $\kappa = 3.0$.
	Those used for (B) correspond to those for curve (B) in 
	Fig. \ref{kekkagraf} when $a=1.0$.
	Likewise, those used for (C) correspond to those for curve (C) in 
	Fig. \ref{kekkagraf}(C) when $a=1.0$.
}
\label{yuudo}
\end{figure}

Next, we demonstrate that the logarithmic marginal likelihood, curve (C) in
Fig. \ref{kekkagraf}, does not have a local maximum and that
the iteratively obtained solution of Eq. (\ref{q}) corresponds to the global maximum. 
Curve (B) in Fig. \ref{yuudo} shows the mean squared error, 
$E_1$, given by Eq.(\ref{Eend}), and curve (C) 
shows the logarithmic marginal likelihood, $\ln(P_{app} (\mbx{\tau}))$,
given by Eq. (\ref{syuhenapp}). 
The horizontal axis denotes $\hat{r}$. 
Curve (B) shows that $E_1$ takes a minimum value, $E_{1}=0.12$, at $\hat{r}=3.2$. 
The parameters used for curve (B) correspond to those for  
curve (B) in Fig. \ref{kekkagraf} when $a=1.0$. 
Curve (C) shows that that the logarithmic marginal likelihood 
takes a maximum value, $E_{1}=0.28$, at $\hat{r}=0.2$. 
The parameters used for curve (C) correspond to those for 
curve (C) in Fig. \ref{kekkagraf} when $a=1.0$. 
The important point is that the logarithmic marginal likelihood, 
curve (C) in Fig. \ref{yuudo}, does not have a local maximum. 
\subsection{Sample of images}
Next, we show the practical significance of these differences
for a typical set of artificial images. 
\begin{figure}[ht]
	\centering
	\includegraphics[width=7.5cm,clip]{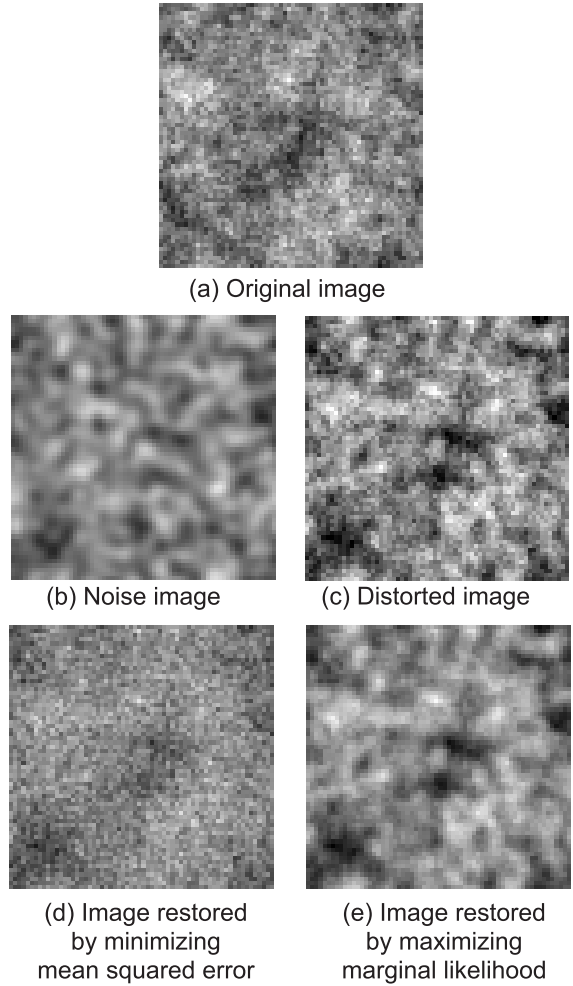}
	\caption{Artificial images. $N=64^2$.
	}
	\label{art}
\end{figure}
Figure \ref{art}(a) shows an original image generated using hyperparameters $\beta=0.5$ and $h=1.0^{-4}$. Figure \ref{art}(c) shows the same image after it was distorted by the noise shown in Fig. \ref{art}(b) ($a=1.0, b=0.75$, and $\kappa=3.0$). 
The mean squared error, $E_2$, between the original and distorted images was 0.57. Figure \ref{art}(d) shows the image after it was restored using the complete restoration model: the generation model and noise model were consistent with the original models, and optimum values were used for the hyperparameters. 
When $a=1.0$, curve (A) in Fig. \ref{kekkagraf} corresponds to Fig. \ref{art}(d),
and the restored mean squared error, $E_1$, is 0.27. 
Figure \ref{art}(e) shows the image after it was restored using an uncorrelated noise model with hyperparameters obtained by maximizing the marginal likelihood. When $a=1.0$, curve (D) in Fig. \ref{kekkagraf} corresponds to Fig. \ref{art}(e), and the restored mean squared error is 0.45. 
The image in Fig. \ref{art}(d) resembles the original one, Fig. \ref{art}(a), while the one in Fig. \ref{art}(e) is similar to the distorted one in Fig. \ref{art}(c) and looks slightly out of focus.

Furthermore, we show the results when natural images were used.
\begin{figure}[ht]
	\centering
	\includegraphics[width=6.74cm,clip]{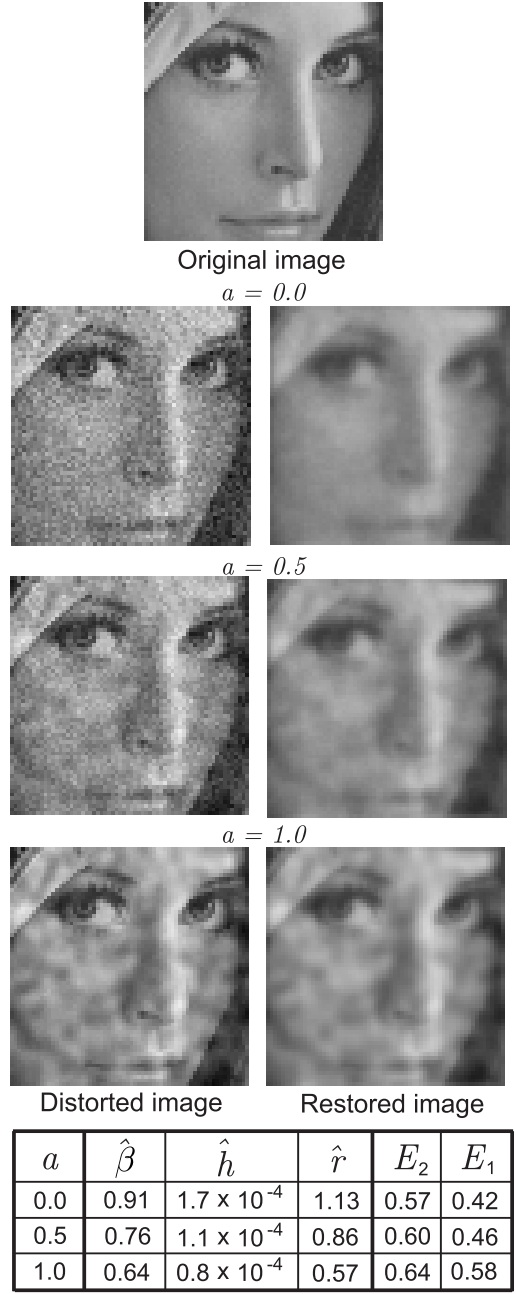}
	\caption{Natural images. $N=64^2$.
	}
	\label{natural}
\end{figure}
Figure \ref{natural} shows the natural images and a table of the estimated parameters. 
We used $b = 0.75$ and $\kappa = 3.0$ for the noise parameters and three values for $a$: 0.0, 0.5, and 1.0. 
The three images on the left are distorted, 
and those on the right are the same images restored using 
$\hat{\beta}$, $\hat{h}$, and $\hat{r}$ estimated using 
the maximized marginal likelihood. 
In this simulation as well, $E_1$ increased with the noise correlation parameter, $a$. 
This tendency is reflected in the three restored images. For $a=1.0$, the lower right image, the noise looks like a stain that was not completely removed. 

\subsection{Discussion}
\begin{figure}[ht]
	\centering
	\includegraphics[width=8cm,clip]{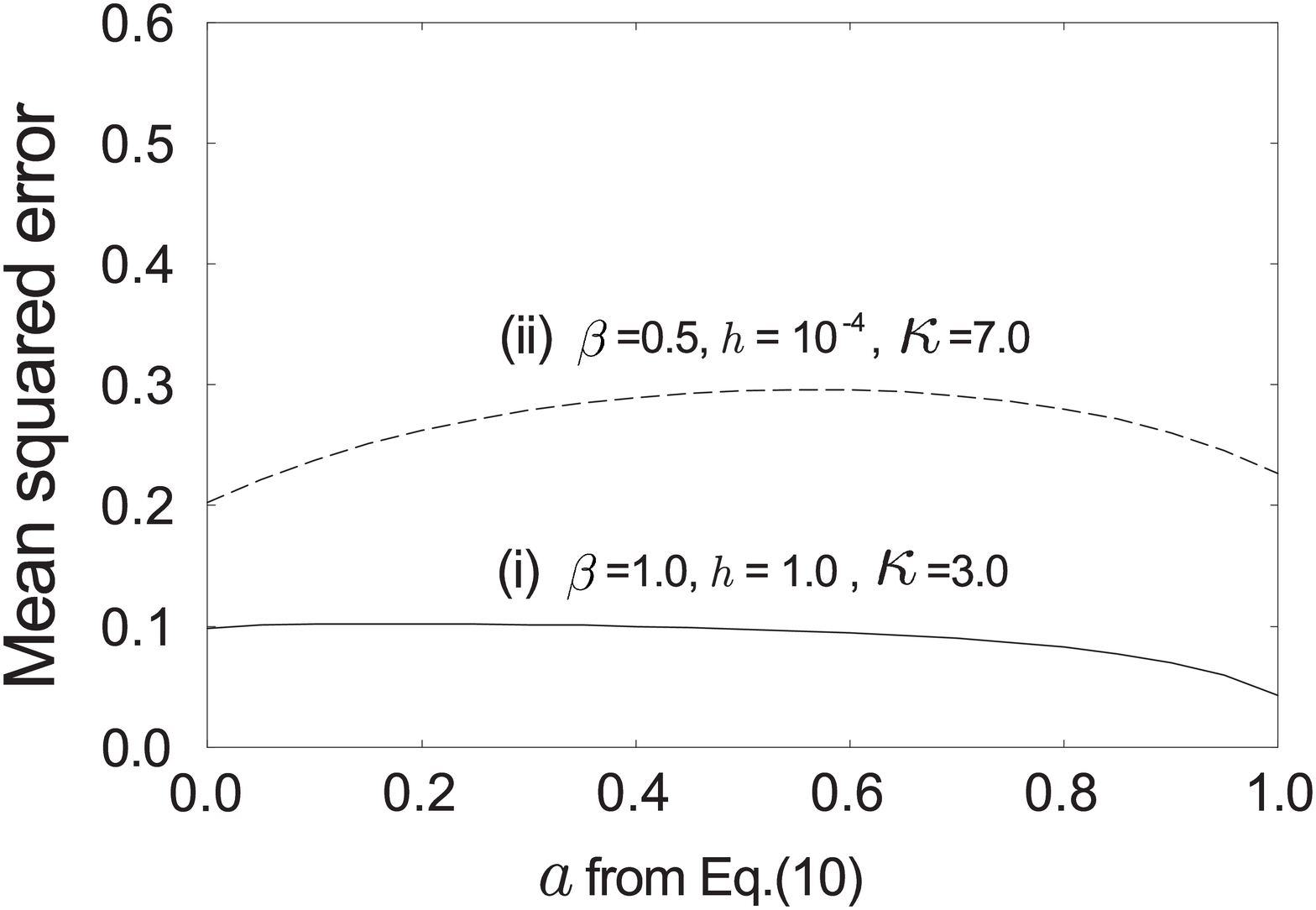}
\caption{
	Mean squared error between original image and restored image. 
	The horizontal axis denotes $a$ from Eq. (\ref{soukan}), and the	vertical axis denotes the mean squared error. 
	As a criterion, the optimal decode given by Eq. (\ref{saitei}) is used. 
	The parameters $b=0.75$, and $N=64^2$ are fixed. 
	For the solid line (i), 
	the other parameters were set as $\beta=1.0, h=1.0$, and $\kappa=3.0$. 
	For the dashed line (ii), 
	the other parameters were set as $\beta=0.5, h=10^{-4}$, and $\kappa=7.0$. 
}
\label{a_eps}
\end{figure}

Here, we consider why line (A) in Fig. \ref{kekkagraf} tends to decrease 
as $a$ approaches $1.0$. This phenomenon appears to be universal, and also occurs with another parameter set as shown in Fig. \ref{a_eps}(ii). 
For reference, we also show line (i) which is based on the parameter set that was used in Fig. \ref{kekkagraf}(A) with only $N$ changed to $N=64^2$.
(Note that line (A) is independent of $N$). 

\begin{figure}[ht]
	\centering
	\includegraphics[width=8cm,clip]{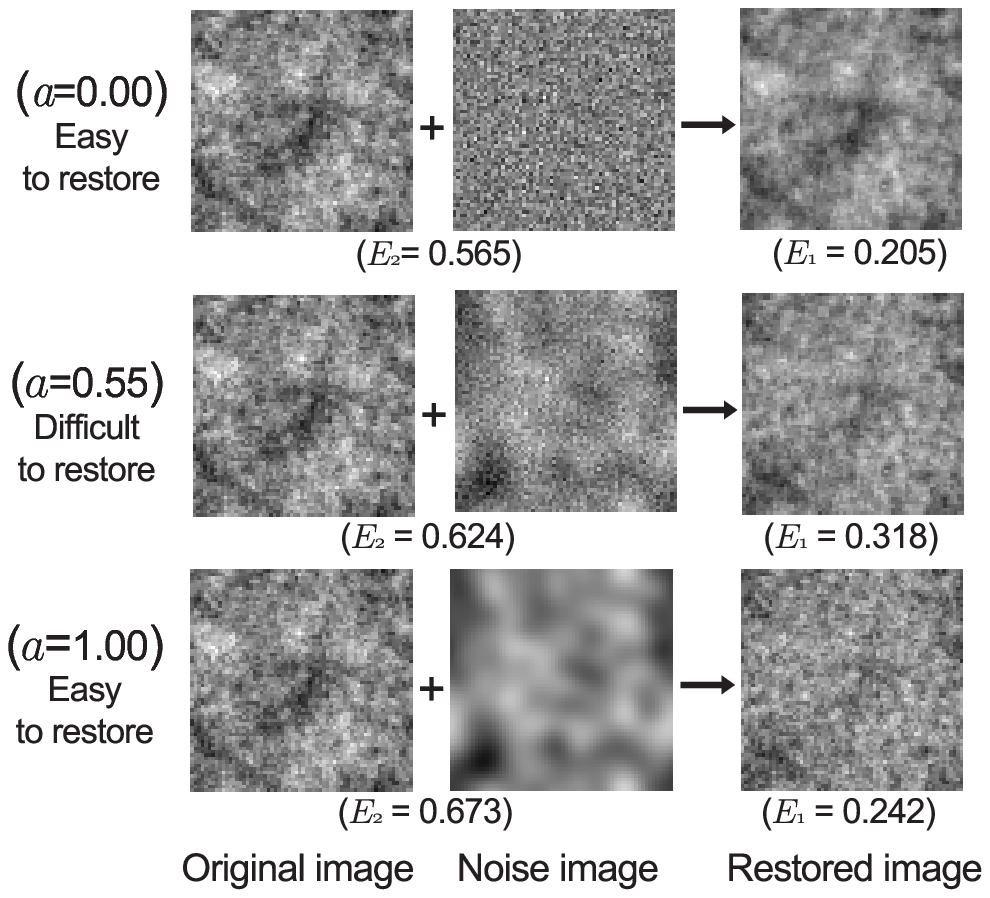}
\caption{
	Typical images about Fig. \ref{a_eps} (ii). $N=64^2$. 
	Parameters for the original images were $\beta=0.5$ and $h=10^{-4}$. The three original prepared images were the same. 
	Parameters for noise images were $\beta=0.5, h=10^{-4}, \kappa=7.0$, and $b=0.75$. Only $a$ was changed. 
	Note that when $a=0.55$, the original and noise images were visually similar, and the restoration quality was lower in this case. 
}
\label{samp_eps}
\end{figure}

For the original image, we used the Gaussian model represented by Eq. (\ref{prior}), 
which has neighboring interactions on each pixel. For the noise image, we used the Gaussian model represented by Eq. (\ref{noize}), 
in which the interaction is Gaussian functionally decreased. 
We expected the difference 
between these two models to become conspicuously large near $a=1.0$, and this difference between the models would enable successful restoration near $a=1.0$. 
Figure \ref{samp_eps} shows the typical images about Fig. \ref{a_eps}(ii). 
The distorted image was most difficult to restore in Fig. \ref{samp_eps} when $a=0.55$, as shown in Fig. \ref{a_eps}(ii).
In this case, the original and the noise images were visually similar; 
more precisely, the correlation lengths of the two images were similar. 
Thus, to successfully restore a distorted image, 
it is desirable that the statistical properties of the original image 
should be different from those of the noise image. 

Here, we summarize our explanation of why line (A) in Fig. \ref{kekkagraf}
decreases near $a=1.0$. 
As $a$ approaches $1.0$, the mean squared error decreases because the difference 
of the statistical properties, especially the correlation length, between the two images enables easier restoration.
Conversely, the mean squared error is greater
at $a=0.55$ on line (ii) in Fig. \ref{a_eps} 
because 
the correlation-length similarity between the images makes it difficult to distinguish between them, 
and that also makes it difficult to restore the distorted image. 
To ensure successful restoration, 
the correlation length of the original image should be different from that of the noise image, which is the case when $a=0.00$ and $a=1.00$ in Fig. \ref{samp_eps}.

\section{Conclusion}
We investigated the use of the Bayesian inference to restore images under conditions of spatially correlated noise. We assumed that both the original image and the noise obeyed a Gaussian distribution composed of translational symmetric matrices. We used the Fourier transformation to diagonalize the covariance matrices, which enabled us to apply various forms of statistical analysis. 
We obtained the expected value of a restored image and obtained 
the optimal analytical hyperparameters by minimizing the mean squared error and used 
 these hyperparameters to determine the generation and noise models. 
Furthermore, we discussed whether 
the conventional spatially uncorrelated noise model could cope with the spatially correlated noise or not. 
We used two methods to estimate the hyperparameters: minimizing the mean squared error and maximizing the marginal likelihood. The difference between the errors obtained using the two methods increased with the noise correlation parameter. The restoration error was larger when we used the hyperparameters obtained using the maximization of the marginal likelihood method. 
Thus, the conventional spatially uncorrelated noise model could not cope with the spatially correlated noise. 

\end{document}